# Site Preferences and "Coloring Problem" in Cu-doped $BiMn_7O_{12}$ Quadruple Perovskite

*Cheng Peng[1&], Mingyu Xu[1&], Yang Zhang[2], Ismail El Baggari[2,3], Jie Li[4], Weiwei Xie[1*]*

1. Department of Chemistry, Michigan State University, East Lansing, MI 48824 USA
2. The Rowland Institute at Harvard, Harvard University, Cambridge, MA 02138 USA
3. Department of Physics, University of British Columbia, Vancouver, BC V6T 1Z4 Canada
4. Department of Earth and Environmental Sciences, University of Michigan, Ann Arbor, MI 48109 USA

[*]Corresponding author: Dr. Weiwei Xie (xieweiwe@msu.edu)

[&]equally contributed

***Abstract***

Lightly Cu-doped $BiMn_{7-x}Cu_xO_{12}$ ($x$ = 0.05, 0.10, and 0.15) was investigated using high-pressure synthesis, single-crystal X-ray diffraction, pair distribution function (PDF) analysis, STEM, magnetic measurements, and first-principles calculations. All compositions retain an average monoclinic $I2/m$ structure, while Cu substitution progressively suppresses the monoclinic distortion and drives the lattice toward a pseudo-cubic metric symmetry. PDF analysis reveals increasing local structural disorder and reduced medium-range coherence with increasing Cu concentration, despite preservation of the overall quadruple-perovskite framework. Single-crystal refinements indicate enhanced electron density at the octahedral Mn B sites, suggesting preferential Cu occupation within the $MnO_6$ network rather than the conventional square-planar sites expected for $Cu^{2+}$. Magnetic measurements reveal two characteristic anomalies near $T_1 \sim$ 100-120 K and $T_2 \sim$ 50-60 K, together with pronounced magnetic irreversibility, field-dependent hysteresis, and unsaturated magnetization. Increasing Cu concentration progressively suppresses the low-temperature magnetic state and weakens the field-induced moment. First-principles calculations favor Cu occupation at the square-planar sites, contrasting with the experimental refinements and highlighting strong competitions among local bonding, short-range disorder, and metastability in this highly frustrated quadruple perovskite system.

## Introduction

The emergence of complex electronic states in correlated oxides is often governed by strong coupling among spin, charge, orbital, and lattice degrees of freedom. Among these systems, A-site-ordered quadruple perovskite manganites, $AA'_3B_4O_{12}$, provide an especially rich platform because the coexistence of square-planar A' sites and corner-sharing octahedral B-site networks generates strong competition among Jahn-Teller distortions, charge/orbital ordering, polar instabilities, and magnetism.[1–5] In particular, $AMn_7O_{12}$ compounds exhibit remarkable electronic tunability through the oxidation state and stereochemical activity of the A-site cation.[6–13] Mixed-valence systems such as $CaMn_7O_{12}$ display coupled charge and orbital ordering with giant spin-driven ferroelectric polarization,[14–17] while $BiMn_7O_{12}$ introduces additional competition through the stereochemically active $Bi^{3+}$ $6s^2$ lone pair, leading to multiple symmetry-lowering structural transitions and polar instabilities.[18–28]

Aliovalent Cu substitution further enriches this landscape by introducing additional charge and orbital degrees of freedom. Previous studies reported incommensurately modulated structures, dipole density waves, and composition-dependent structural transitions in $BiMn_{7-x}Cu_xO_{12}$, yet the crystallographic role of Cu substitution remains unresolved.[29,30] This problem is closely connected to the "coloring problem" in solids which describes how chemically similar atoms distribute among multiple crystallographically inequivalent sites in complex solids. In Cu-doped $BiMn_7O_{12}$, the coexistence of several inequivalent Jahn-Teller-active Mn sites, combined with strong local distortions and high-pressure metastability, creates an ideal platform for investigating how local bonding, electronic instability, and synthesis conditions determine cation distribution in correlated oxides synthesized under extreme conditions.

Here, we investigate Cu-doped $BiMn_7O_{12}$ using single-crystal X-ray diffraction, scanning transmission electron microscopy (STEM), magnetic measurements, and total-energy calculations. Contrary to the conventional expectation that Jahn-Teller-active $Cu^{2+}$ ions preferentially occupy the square-planar A' site, our structural refinements reveal that Cu predominantly occupies the distorted octahedral B sites within the quadruple perovskite framework. This site preference fundamentally changes the understanding of the structural evolution in $BiMn_{7-x}Cu_xO_{12}$, indicating that Cu substitution directly perturbs the Mn-O octahedral network rather than primarily modifying the square-planar A' sublattice. Despite the retention of an average monoclinic *I*2/*m* structure in the single-crystal refinements, STEM imaging and total-energy calculations further suggest that Cu site occupancy is highly sensitive to local structural environments and synthesis conditions. These results demonstrate that even subtle variations in local structural topology and cation distribution can significantly influence the balance among orbital, polar, and magnetic instabilities in quadruple perovskite manganites.

## Experimental Methods and Calculations

**High-Pressure Synthesis:** $BiMn_{7-x}Cu_xO_{12}$ single crystals were synthesized under high-pressure and high-temperature conditions using the Walker-type multi-anvil press at the University of Michigan.[31] Stoichiometric mixtures of $Bi_2O_3$ (bismuth oxide, 99%, thermo scientific), $Mn_2O_3$, $MnO_2$ (manganese oxide, 99%, thermo scientific), and CuO (copper oxide, 99%, thermo scientific) were used as starting materials with a nominal precursor ratio of 1: (7-2*x*): 2*x*: 2*x*. Single-phase $Mn_2O_3$ was prepared by heating $MnO_2$ in air at 650 °C for 24 hours. The precursor powders were thoroughly ground and sealed in an Au capsule. The pressure assembly consisted of a Ceramacast 584 octahedral pressure medium, a Re heater, and Toshiba Tungaloy tungsten-carbide anvils with 8-mm truncation edge length.[32] The sample was compressed to 6.5 GPa at room temperature over 12 h, heated to 1100 °C, and held at this temperature for 1 h. The sample was then quenched to room temperature before slow decompression to ambient pressure. Single crystals of $BiMn_{7-x}Cu_xO_{12}$, with typical dimensions of approximately 0.3 mm, were obtained within a polycrystalline $BiMn_{7-x}Cu_xO_{12}$ matrix.

**Ambient-Pressure Single-Crystal X-ray Diffraction:** Single-crystal X-ray diffraction measurements were performed on selected $BiMn_{7-x}Cu_xO_{12}$ crystals mounted on nylon loops with Paratone oil. Data were collected using a Rigaku XtaLAB Synergy Dualflex diffractometer equipped with a HyPix detector. Measurements were conducted between 80 and 300 K at ambient pressure. Diffraction data were collected using $\omega$ scans with Mo $K_\alpha$ radiation ($\lambda$= 0.71073Å) generated from a micro-focus sealed tube operated at 50 kV and 1 mA. Data collection strategies, including the number of runs and images, were automatically determined using CrysAlisPro software. Data reduction included Lorentz and polarization corrections, numerical absorption correction using Gaussian integration over a multifaceted crystal model,[33] and empirical spherical harmonics correction using the SCALE3 ABSPACK algorithm.[34] The crystal structures were solved and refined using the Bruker SHELXTL package.[35,36] The resulting crystallographic parameters are summarized in **Tables 1** and **2**.

**Pair Distribution Function Analysis:** Pair distribution function (PDF) analysis was performed to qualitatively evaluate the local structure of $BiMn_{7-x}Cu_xO_{12}$. The diffraction data used for PDF conversion was obtained from single-crystal X-ray diffraction frames that were converted into powder-like diffraction profiles. The total scattering intensity was processed and Fourier transformed into real space to obtain the pair distribution function, $G(r)$.[37,38]

$$G(r) = \frac{2}{\pi}\int_{Q_{\min}}^{Q_{\max}} Q[S(Q)-1]\sin(Qr)dQ$$

Because of the limited $Q_{max}$ accessible from laboratory X-ray data, the resulting PDFs have limited real-space resolution and may contain termination ripples. Therefore, the PDF results were used only for qualitative comparison rather than quantitative refinement. Simulated PDFs were generated from the SCXRD-derived crystallographic models using the corresponding CIF files as input. The same $Q$-range and damping/broadening parameters were used for simulated and experimental PDFs to enable direct comparison.

**Magnetization and Electrical Transport Measurements:** Temperature- and magnetic-field-dependent magnetization measurements were performed using a Quantum Design Magnetic Property Measurement System (MPMS3). Electrical resistance measurements were carried out using a Quantum Design Physical Property Measurement System (PPMS DynaCool).

**Scanning Transmission Electron Microscopy:** Scanning transmission electron microscopy (STEM) experiments were performed using an aberration-corrected Thermo Fisher Scientific Themis Z G3 microscope operated at 200 kV.

**First-Principles Calculations:** To further evaluate the site preference of Cu in $BiMn_{7-x}Cu_xO_{12}$, the first-principles calculations were performed using Quantum ESPRESSO v7.4.1.[39,40] The calculations utilized projector augmented-wave (PAW) pseudopotentials in conjunction with the Perdew-Burke-Ernzerhof (PBE) exchange-correlation functional.[41,42] A wavefunction cutoff energy of 100 Ry and a charge density cutoff set to ten times this value were employed. A 6×6×6 Monkhorst-Pack k-points mesh was employed in the reciprocal space.[43] Convergence tests were conducted to ensure the change of total energy was smaller than 1 meV/atom. The Davidson diagonalization algorithm[44] was applied, and the convergence threshold for self-consistency was set to $10^{-9}$ Ry. The conventional $Bi_2Mn_{14}O_{24}$ unit cell, containing 40 atoms and corresponding to two primitive cells, was used as the parent structure. To model the dilute Cu-doping limit, one Mn atom was replaced by one Cu atom in each calculation. Cu substitution at different crystallographic Mn sites was then compared by evaluating the corresponding total energy.

Three computational models were considered. First, spin-polarized calculations were performed assuming a ferromagnetic configuration, with full relaxation of the atomic positions while keeping the lattice parameters fixed. Second, fixed-geometry calculations were carried out using the same ferromagnetic configuration without relaxing the atomic positions, allowing comparison of the intrinsic energetic preference of Cu at different sites in the experimentally refined structure. Third, DFT+ U calculations were performed to account for electron correlation effects on Mn 3*d* states, using an effective Hubbard parameter of $U_{eff} = 4$ eV for Mn 3*d* orbitals.

For all models, consistent computational parameters were used to ensure direct comparison among different Cu-substitution configurations. The relative total energies were used to determine the preferred Cu occupation site and to evaluate the energetic competition between Cu substitutions at the square-planar A’ site and the distorted octahedral B sites.

## Results and Discussion

**Tables 1** and **2** summarize the crystallographic refinement parameters and atomic coordinates of $BiMn_{7-x}Cu_xO_{12}$ obtained from ambient-pressure single-crystal X-ray diffraction measurements at room temperature. All compositions investigated were successfully refined in the monoclinic $I2/m$ space group, indicating that the average crystal structure remains monoclinically distorted across the lightly Cu-doped region. The diffraction data was collected from single crystals with typical dimensions of approximately 0.1 mm extracted from the high-pressure synthesized polycrystalline matrix. Multiple crystals from the same synthesis batch were independently measured and refined, yielding highly consistent lattice parameters and structural solutions, supporting the reproducibility and phase homogeneity of the synthesized samples. The refined lattice parameters show only subtle changes with increasing Cu concentration, suggesting that Cu substitution does not induce a major crystallographic reconstruction within the investigated doping range. Instead, the structural evolution primarily involves modest distortions of the existing quadruple perovskite framework. This behavior contrasts with the strong composition-dependent structural complexity previously reported for $BiMn_{7-x}Cu_xO_{12}$, where multiple symmetry-lowering transitions and incommensurately modulated phases were proposed upon cooling.[29] In particular, previous synchrotron and neutron powder diffraction studies suggested that lightly doped compositions ($x$= 0.05 and 0.10) evolve from the monoclinic $I2/m$ phase into modulated $R\bar{1}(\alpha\beta\gamma)^0$ and polar $R3(00\gamma)t$ phases at lower temperatures.[29] By contrast, the present single-crystal refinements consistently support an average monoclinic $I2/m$ structure at room temperature for all measured compositions, indicating that the long-range average structure remains comparatively robust despite the presence of local disorder and competing structural instabilities. According to the single crystal XRD refinement, Cu substitution progressively suppresses the monoclinic distortion: the lattice parameters converge, the unit-cell volume decreases, and the monoclinic angle β approaches 90°. This indicates that Cu doping drives the structure toward a more metrically pseudo-cubic framework without producing a true symmetry change detectable. The $x$ = 0.15 composition exhibits the strongest pseudosymmetry, but the reflection statistics still favor monoclinic structure over higher-symmetry models. Relatively large $R_{int}$ values, ranging from approximately 0.09 to 0.26, were observed in the present refinements. These elevated values are attributed to the presence of multiple crystallographic domains, local disorder, and strain effects within the crystals, which are commonly encountered in high-pressure synthesized complex oxides. The relatively large residual electron density peaks and holes observed in the single-crystal refinements likely originate from the coexistence of local structural disorder, crystallographic domain formation, and competing Cu/Mn site occupancies within the metastable high-pressure phase. Such behavior is consistent with the large $R_{int}$ values, which indicate substantial local structural inhomogeneity despite retention of the average monoclinic *I*2/*m* symmetry. In addition, previous studies reported incommensurately modulated structures in lightly Cu-doped $BiMn_7O_{12}$, suggesting that weak local symmetry lowering or short-range modulation may remain partially unresolved in the present average structural model. Despite these challenges, the final refinement quality remains relatively acceptable, with $R_1$ values on the order of ~0.1 and goodness-of-fit (GOF) values close to unity (~1.0-1.1). These values indicate that the refined structural models provide a reliable description of the diffraction data and capture the essential crystallographic features of the system.

Importantly, the present single-crystal refinements reveal that Cu preferentially occupies the distorted octahedral B sites rather than the square-planar A' sites typically favored by $Cu^{2+}$ ions in many $AA'_3B_4O_{12}$ quadruple perovskites.[45,29] This observation differs from the conventional expectation for square-planar coordinated $Cu^{2+}$ and suggests that the energetic balance between Jahn-Teller stabilization and local lattice strain in $BiMn_{7-x}Cu_xO_{12}$ is unusually delicate. Consequently, Cu substitution directly perturbs the Mn-O octahedral network and may significantly influence the orbital ordering tendencies, local structural distortions, and magnetic exchange interactions within the system.

**Table 1.** Crystal structure and refinement data for $BiMn_{7-x}Cu_xO_{12}$. Values in parentheses represent estimated standard deviations from the refinements.

| | $BiMn_7O_{12}$ | $BiMn_{6.95}Cu_{0.05}O_{12}$ | $BiMn_{6.90}Cu_{0.10}O_{12}$ | $BiMn_{6.85}Cu_{0.15}O_{12}$ |
|---|---|---|---|---|
| Temperature(K) | **295** | **295** | **295** | **295** |
| Space Group | ***I*2/*m*** | ***I*2/*m*** | ***I*2/*m*** | ***I*2/*m*** |
| Unit Cell dimensions | ***a*** **= 7.5065(5) Å**<br>***b*** **= 7.3904(6) Å**<br>***c*** **= 7.5084(5) Å**<br>***β*** **= 90.870(7)°** | ***a*** **= 7.5020(5) Å**<br>***b*** **= 7.3786(6) Å**<br>***c*** **= 7.5216(5) Å**<br>***β*** **= 91.114(6)°** | ***a*** **= 7.4692(10) Å**<br>***b*** **= 7.4248(11) Å**<br>***c*** **= 7.4833(10) Å**<br>***β*** **= 90.612(6)°** | ***a*** **= 7.4490(12) Å**<br>***b*** **= 7.4494(12) Å**<br>***c*** **= 7.4597(15) Å**<br>**β = 90.472(16)** |
| Volume | **416.49(5) Å$^3$** | **416.27(5)** | **414.98(10)** | **413.93(13)** |
| Z | **2** | **2** | **2** | **2** |
| Density (calculated) | **6.264** | **6.271** | **6.294** | **5.978** |
| Absorption coefficient(μ/mm$^{-1}$) | **31.372** | **31.440** | **31.590** | **27.470** |
| F (000) | **708** | **708** | **709** | **676** |

| 2θ range | **7.618 to 81.186** | **7.598 to 82.636** | **7.668 to 82.732** | **7.7 to 95.828** |
|---|---|---|---|---|
| Reflections collected | **14942** | **11666** | **4768** | **10130** |
| Independent reflections | **1414 [$R_{int}$ = 0.142]** | **1435 [$R_{int}$ = 0.183]** | **1415 [$R_{int}$ = 0.0888]** | **1503 [$R_{int}$ = 0.2556]** |
| Data/restraints/parameters | **1414/0/59** | **1435/0/60** | **1415/0/59** | **1503/0/59** |
| Final $R$ indices | **$R_1$ (I>2σ(I)) = 0.0864; $wR_2$ (I > 2σ(I)) = 0.2180 $R_1$ (all) = 0.0887; $wR_2$ (all) = 0.2210** | **$R_1$ (I>2σ(I)) = 0.0917; $wR_2$ (I > 2σ(I)) = 0.2345 $R_1$ (all) = 0.0972; $wR_2$ (all) = 0.2404** | **$R_1$ (I>2σ(I)) = 0.0991; $wR_2$ (I > 2σ(I)) = 0.2527 $R_1$ (all) = 0.1092; $wR_2$ (all) = 0.2619** | **$R_1$ (I>2σ(I)) = 0.1201; $wR_2$ (I > 2σ(I)) = 0.3008 $R_1$ (all) = 0.1555; $wR_2$ (all) = 0.3281** |
| Largest diff. peak and hole | **+8.79 $e^-/Å^3$ and -7.63 $e^-/Å^3$** | **+10.67 $e^-/Å^3$ and -13.19 $e^-/Å^3$** | **+10.79 $e^-/Å^3$ and -4.43$e^-/Å^3$** | **+8.97 $e^-/Å^3$ and -3.95 $e^-/Å^3$** |
| Goodness-of-fit on $F^2$ | **1.078** | **1.036** | **1.083** | **1.117** |

**Table 2.** Atomic coordinates and equivalent isotropic atomic displacement parameters ($Å^2$) for $BiMn_{7-x}Cu_xO_{12}$ at room temperature. The Bi-site occupancy for the $x$ = 0.15 crystal was refined to 0.8, suggesting possible Bi deficiency or local disorder in the measured crystal.

| | **$T$ (K)** | **Atom** | **Wyck.** | ***x*** | ***y*** | ***z*** | **Occ.** | **$U_{eq}$** |
|---|---|---|---|---|---|---|---|---|
| **$BiMn_7O_{12}$** | **295** | **Bi** | 2*c* | 0 | 0 | 1/2 | 1 | 0.015(19) |
| | | **$Mn_{B1}$** | 4*e* | 1/4 | 1/4 | 0 | 1 | 0.033(7) |
| | | **$Mn_{B2}$** | 4*f* | 1/4 | 1/4 | 1/2 | 1 | 0.022(5) |
| | | **$Mn_{A1}$** | 2*b* | 0 | 1/2 | 0 | 1 | 0.036(11) |
| | | **$Mn_{A2}$** | 2*a* | 0 | 0 | 0 | 1 | 0.030(9) |
| | | **$Mn_{A3}$** | 2*d* | 0 | 1/2 | 1/2 | 1 | 0.028(7) |
| | | **O1** | 4*i* | 0.173(5) | 0 | 0.354(8) | 1 | 0.089(12) |
| | | **O2** | 4*i* | 0.182(5) | 0 | 0.001(8) | 1 | 0.083(11) |
| | | **O3** | 8*j* | 0.186(5) | 0.321(3) | 0.178(6) | 1 | 0.099(10) |
| | | **O4** | 8*j* | 0.010(4) | 0.316(4) | 0.336(5) | 1 | 0.081(7) |
| **$BiMn_{6.95}Cu_{0.05}O_{12}$** | **295** | **Bi** | 2*c* | 0 | 0 | 1/2 | 1 | 0.015(2) |
| | | **$Mn_{B1}$** | 4*e* | 1/4 | 1/4 | 0 | 0.99(1) | 0.031(8) |
| | | **$Cu_{B1}$** | 4*e* | 1/4 | 1/4 | 0 | 0.01(1) | 0.031(8) |
| | | **$Mn_{B2}$** | 4*f* | 1/4 | 1/4 | 1/2 | 0.99(1) | 0.014(4) |
| | | **$Cu_{B2}$** | 4*f* | 1/4 | 1/4 | 1/2 | 0.01(1) | 0.014(4) |
| | | **$Mn_{A1}$** | 2*b* | 0 | 1/2 | 0 | 1 | 0.031(11) |
| | | **$Mn_{A2}$** | 2*a* | 0 | 0 | 0 | 1 | 0.022(7) |
| | | **$Mn_{A3}$** | 2*d* | 0 | 1/2 | 1/2 | 1 | 0.019(6) |
| | | **O1** | 4*i* | 0.163(5) | 0 | 0.356(6) | 1 | 0.081(13) |
| | | **O2** | 4*i* | 0.182(4) | 0 | 0.004(6) | 1 | 0.072(11) |
| | | **O3** | 8*j* | 0.185(5) | 0.322(2) | 0.174(6) | 1 | 0.103(12) |
| | | **O4** | 8*j* | 0.016(3) | 0.312(2) | 0.340(4) | 1 | 0.062(6) |
| **$BiMn_{6.90}Cu_{0.10}O_{12}$** | **295** | **Bi** | 2*c* | 0 | 0 | 1/2 | 1 | 0.020(3) |
| | | **$Mn_{B1}$** | 4*e* | 1/4 | 1/4 | 0 | 0.97(1) | 0.029(6) |
| | | **$Cu_{B1}$** | 4*e* | 1/4 | 1/4 | 0 | 0.03(1) | 0.029(6) |
| | | **$Mn_{B2}$** | 4*f* | 1/4 | 1/4 | 1/2 | 0.97(1) | 0.021(5) |
| | | **$Cu_{B2}$** | 4*f* | 1/4 | 1/4 | 1/2 | 0.03(1) | 0.021(5) |
| | | **$Mn_{A1}$** | 2*b* | 0 | 1/2 | 0 | 1 | 0.030(8) |
| | | **Mn$_{A2}$** | 2*a* | 0 | 0 | 0 | 1 | 0.027(7) |
| | | **Mn$_{A3}$** | 2*d* | 0 | 1/2 | 1/2 | 1 | 0.025(7) |
| | | **O1** | 4*i* | 0.172(5) | 0 | 0.351(8) | 1 | 0.082(11) |
| | | **O2** | 4*i* | 0.180(5) | 0 | 0.003(8) | 1 | 0.078(10) |
| | | **O3** | 8*j* | 0.178(5) | 0.321(3) | 0.174(6) | 1 | 0.082(9) |
| | | **O4** | 8*j* | 0.006(4) | 0.314(4) | 0.331(5) | 1 | 0.080(7) |
| **$BiMn_{6.85}Cu_{0.15}O_{12}$** | **295** | **Bi** | 2*c* | 0 | 0 | 1/2 | 0.80(2) | 0.018(3) |
| | | **Mn$_{B1}$** | 4*e* | 1/4 | 1/4 | 0 | 0.96(1) | 0.026(6) |
| | | **Cu$_{B1}$** | 4*e* | 1/4 | 1/4 | 0 | 0.04(1) | 0.026(6) |
| | | **Mn$_{B2}$** | 4*f* | 1/4 | 1/4 | 1/2 | 0.96(1) | 0.021(5) |
| | | **Cu$_{B2}$** | 4*f* | 1/4 | 1/4 | 1/2 | 0.04(1) | 0.021(5) |
| | | **Mn$_{A1}$** | 2*b* | 0 | 1/2 | 0 | 1 | 0.023(7) |
| | | **Mn$_{A2}$** | 2*a* | 0 | 0 | 0 | 1 | 0.024(7) |
| | | **Mn$_{A3}$** | 2*d* | 0 | 1/2 | 1/2 | 1 | 0.023(7) |
| | | **O1** | 4*i* | 0.183(5) | 1/2 | 0.000(7) | 1 | 0.080(9) |
| | | **O2** | 4*i* | 0.176(6) | 1/2 | 0.344(8) | 1 | 0.085(7) |
| | | **O3** | 8*j* | 0.182(3) | 0.319(4) | 0.681(5) | 1 | 0.087(11) |
| | | **O4** | 8*j* | 0.008(3) | 0.175(3) | 0.829(5) | 1 | 0.083(7) |

**Figure 1** summarizes the structural evolution of $BiMn_{7-x}Cu_xO_{12}$ as a function of Cu substitution. The parent compound crystallizes in the monoclinic *I*2/*m* space group, characteristic of A-site-ordered quadruple perovskites with the general formula $AA'_3B_4O_{12}$. As illustrated in **Figure 1A**, the structure consists of a three-dimensional framework of corner-sharing $MnO_6$ octahedra, while Bi occupies the 12-fold coordinated A site. Owing to the strong octahedral tilting and Jahn-Teller distortions, five crystallographically inequivalent Mn sites are present. $Mn_{A1}$, $Mn_{A2}$, and $Mn_{A3}$ correspond to the square-planar coordinated A' sites, whereas $Mn_{B1}$ and $Mn_{B2}$ occupy the conventional octahedral B sites within the perovskite framework. Systematic evolution of the lattice parameters is observed with increasing Cu concentration (**Figure 1B**). The *a* and *c* lattice parameters gradually decrease, while the *b*parameter increases slightly, leading to a progressive convergence of the lattice constants. Simultaneously, the monoclinic angle β decreases toward 90°, indicating suppression of the monoclinic distortion and a gradual evolution toward a more metrically cubic framework. This behavior suggests that Cu substitution partially relieves the structural distortion associated with the Mn network and stabilizes a more symmetric average structure. Consistent with this trend, the unit-cell volume decreases monotonically with increasing Cu concentration (**Figure 1C**), indicating an overall lattice contraction upon Cu incorporation. The continuous reduction in cell volume suggests that Cu substitution modifies the local bonding environment and electronic structure of the Mn-O framework. Such behavior likely reflects the combined effects of differences in ionic radii, covalency, and local Jahn-Teller stabilization between Mn and Cu ions. More importantly, the site-dependent electron density analysis obtained from single-crystal X-ray diffraction refinements reveals a highly non-random distribution of Cu among the five inequivalent Mn sites (**Figure 1D**). The relative electron density increases significantly at the $Mn_{B1}$ and $Mn_{B2}$ octahedral sites with increasing Cu concentration, while the electron density at the square-planar $Mn_A$ sites either remains nearly unchanged or decreases slightly. In particular, the $Mn_{A3}$ site shows the strongest reduction in relative electron density upon Cu substitution. These results indicate that Cu preferentially occupies the distorted octahedral B sites rather than the square-planar A' sites typically favored by $Cu^{2+}$ ions in many $AA'_3B_4O_{12}$ quadruple perovskites. This unconventional site preference is highly significant because it directly perturbs the $MnO_6$ octahedral network responsible for orbital ordering and magnetic exchange interactions. The preferential occupation of the B sublattice by Cu suggests that the energetic balance between Jahn-Teller distortions, local lattice strain, and Bi lone-pair-driven polar instabilities in $BiMn_{7-x}Cu_xO_{12}$ is unusually delicate. Consequently, even light Cu substitution can substantially modify the local electronic structure and magnetostructural coupling behavior without inducing a major long-range crystallographic reconstruction.

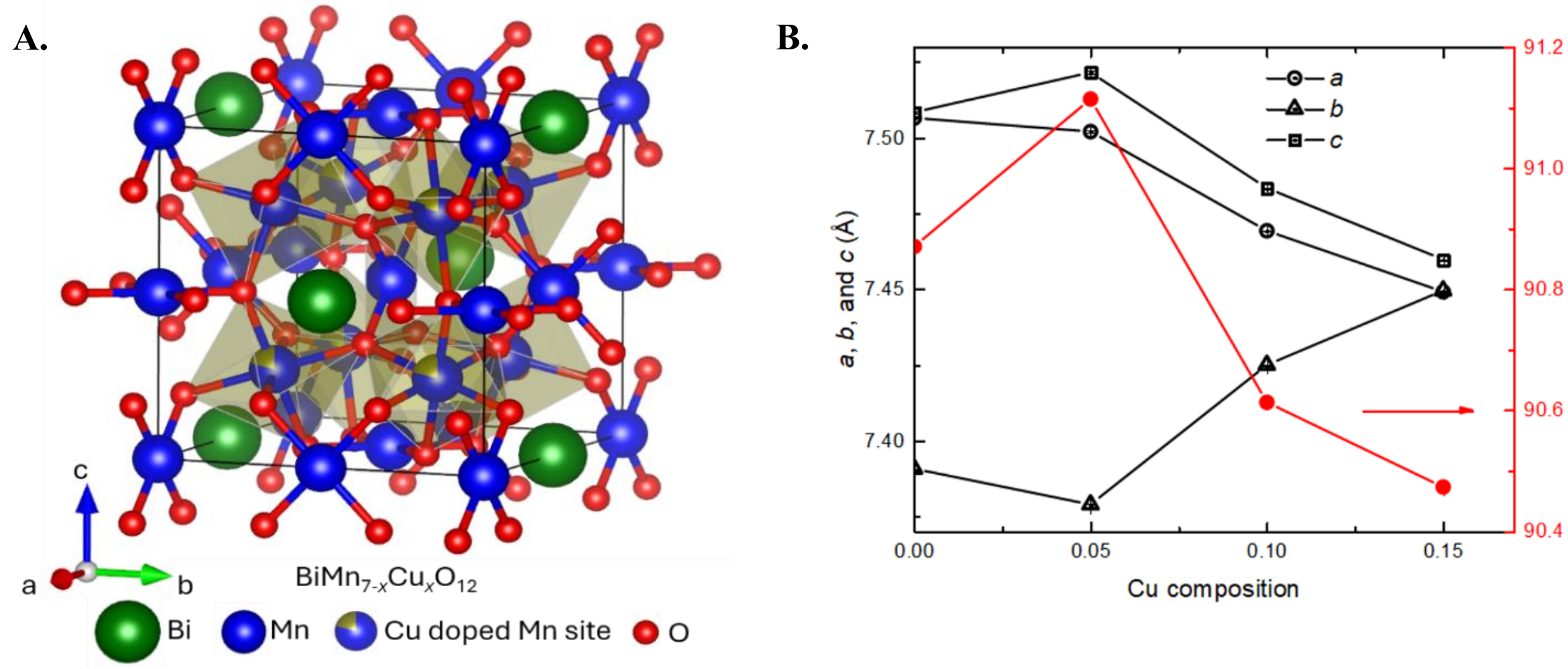

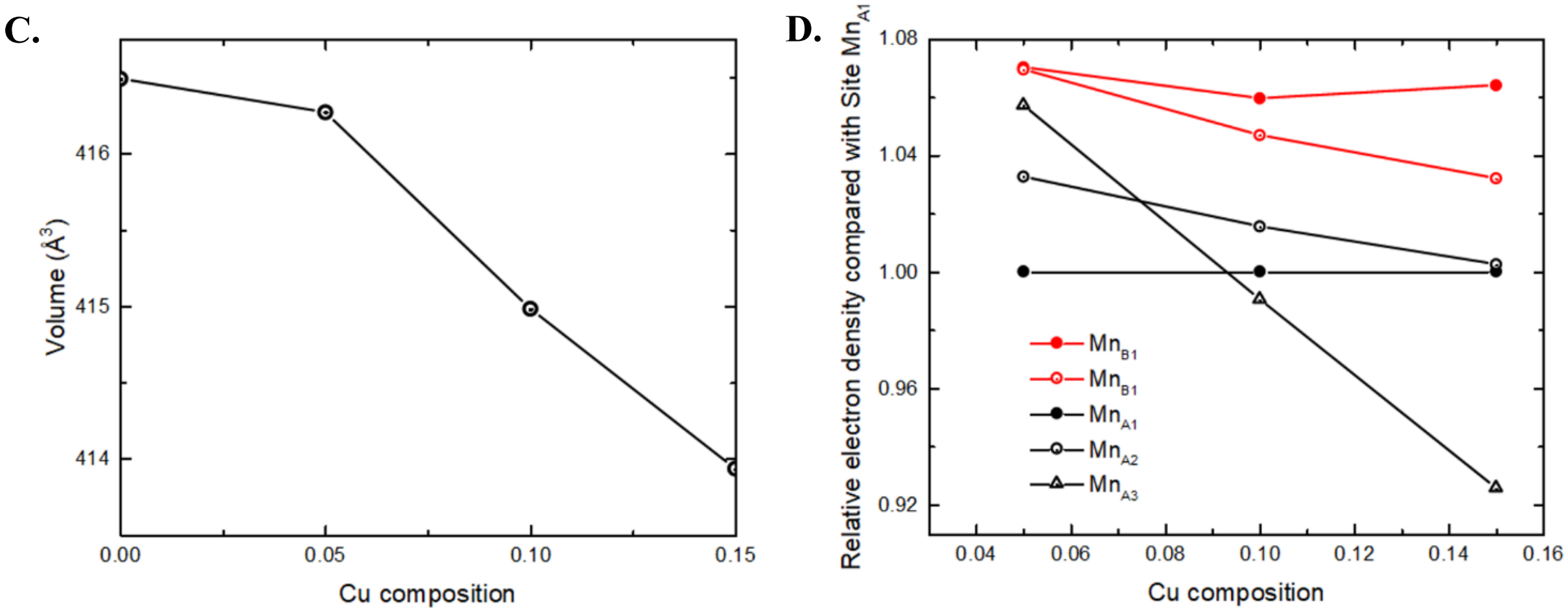

**Figure 1. Structural evolution and Cu site preference in $BiMn_{7-x}Cu_xO_{12}$. (A)** Crystal structure of $BiMn_{7-x}Cu_xO_{12}$ refined in the monoclinic *I*2/*m* space group. **(B)** Evolution of lattice parameter (*a*, *b*, *c*), together with the monoclinic angle β, as a function of Cu composition. **(C)** Unit-cell volume as a function of Cu concentration, showing progressive lattice contraction upon Cu substitution. **(D)** Relative electron density at the five inequivalent Mn sites, normalized to the $Mn_{A1}$ site, revealing preferential Cu occupation at the octahedral $Mn_B$ sites.

**Figure 2** presents the temperature dependence of the lattice parameters and unit-cell volume of the parent $BiMn_7O_{12}$ obtained from single-crystal X-ray diffraction measurements during both warming and cooling cycles. Throughout the investigated temperature range, the crystal structure remains well described by the monoclinic *I*2/*m* space group, and no additional reflections or symmetry changes associated with a long-range structural phase transition were observed. As shown in **Figure 2A**, the lattice parameters *a, b, c*, together with the monoclinic angle β, exhibit only subtle temperature-dependent variations. The *a* and *c* lattice parameters remain relatively stable with weak anisotropic thermal evolution, while the *b* parameter shows a slight increase upon warming. The monoclinic angle β remains close to 91° over the entire temperature range, indicating that the monoclinic distortion is preserved and that no transition toward a higher-symmetry structure occurs. The absence of abrupt anomalies or discontinuities in the lattice parameters further supports the lack of a symmetry-breaking structural transition within the measured temperature window. The unit-cell volume shown in **Figure 2B** exhibits a modest but reproducible hysteresis between warming and cooling cycles. Rather than arising solely from systematic experimental uncertainty, this hysteretic behavior likely reflects subtle metastability, local strain relaxation, or slow structural dynamics associated with competing lattice distortions in$BiMn_7O_{12}$. Similar behavior is consistent with the delicate balance among Jahn-Teller distortions, octahedral tilting, and $Bi^{3+}$ lone-pair-driven polar instabilities known in quadruple perovskite manganites.[25,28] Importantly, no abrupt volume collapse or discontinuity is observed, indicating that the hysteresis originates from subtle local structural rearrangements rather than a first-order crystallographic phase transition. These results demonstrate that $BiMn_7O_{12}$ maintains a robust monoclinic *I*2/*m* average structure across the investigated temperature range while exhibiting weak thermally driven lattice adjustments and metastable distortion dynamics. The coexistence of structural robustness and subtle hysteretic lattice responses further suggests strong coupling between local structural distortions and the competing electronic instabilities in this system. This temperature-dependent structural stability provides an important reference point for understanding the composition-dependent evolution of the Cu-doped $BiMn_{7-x}Cu_xO_{12}$ series, where Cu substitution progressively modifies the monoclinic distortion without producing an abrupt long-range symmetry change.

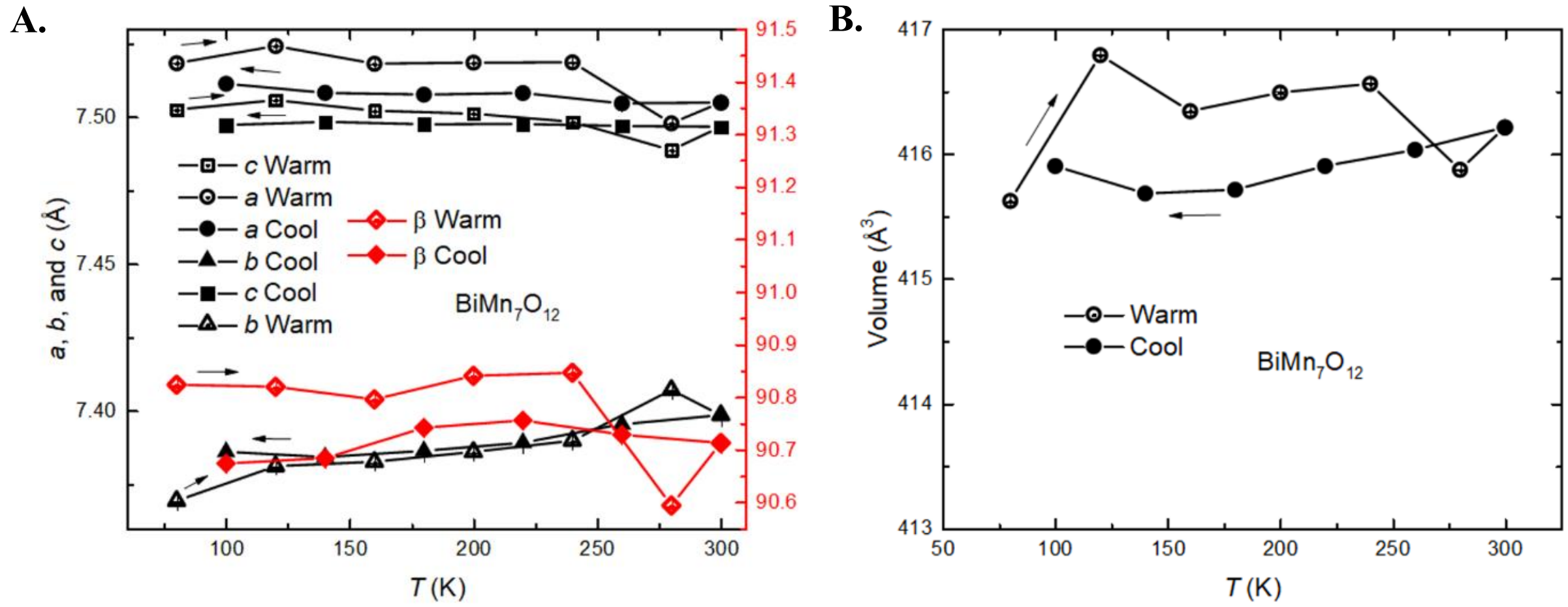

**Figure 2. Temperature-dependent structural evolution of $BiMn_7O_{12}$ (A)** Temperature dependence of the lattice parameters (*a*, *b*, *c*), together with the monoclinic angle β, measured during warming and cooling cycles. The lattice parameters exhibit subtle anisotropic thermal evolution while preserving the monoclinic *I*2/*m* structure over the investigated temperature range. **(B)** Temperature dependence of the unit-cell volume of $BiMn_7O_{12}$, showing a small but reproducible hysteresis between warming and cooling cycles without any abrupt volume discontinuity associated with a symmetry-breaking structural transition.

Pair distribution function (PDF) analysis was performed to investigate the local structural evolution of $BiMn_{7-x}Cu_xO_{12}$ as a function of Cu substitution shown in **Figure 3**. (**Figures S1-S4** present the total scattering structure factor, $F(Q)$ and corresponding pair distribution function, $G(r)$ for $x$ = 0, 0.05, 0.10, and 0.15, respectively.) Overall, all compositions exhibit very similar PDF peak positions throughout the measured *r*-range, indicating that the fundamental quadruple perovskite framework remains preserved upon Cu substitution. The absence of major peak shifts or new correlation features suggests that Cu incorporation does not induce a large-scale reconstruction of the local atomic network or a symmetry-breaking structural transition. This observation is fully consistent with the single-crystal X-ray diffraction refinements, which indicate retention of the monoclinic $I2/m$ average structure for all investigated compositions. Despite the overall similarity, systematic changes are clearly observed in both the $F(Q)$ and $G(r)$ profiles with increasing Cu concentration. In the parent compound $BiMn_7O_{12}$ (**Figure S1**), the PDF oscillations remain relatively sharp and well defined over a broad *r*-range, indicating comparatively strong medium-range structural coherence. Upon Cu substitution (**Figures S2-S4**), gradual broadening and damping of the PDF oscillations become increasingly pronounced, particularly in the intermediate- and long-range regions ($r > 10$ Å). The suppression of the oscillation amplitude and changes in relative peak intensity indicate increasing local structural disorder and reduced medium-range coherence associated with Cu incorporation. The evolution of the total scattering structure factor $F(Q)$ further supports this interpretation. Although the primary scattering peaks remain largely preserved across the doping series, subtle variations in peak intensity and diffuse scattering background are observed with increasing Cu concentration, indicating enhanced local disorder within the Mn-O framework. Because PDF analysis is highly sensitive to local bonding environments, these changes likely originate from local perturbations of Mn-O and Cu-O bond distances, octahedral tilting, and modifications of Jahn–Teller distortions induced by Cu substitution. These local structural modifications are particularly significant when considered together with the unconventional Cu site preference identified from the single-crystal refinements. Rather than occupying the square-planar A' sites commonly favored by $Cu^{2+}$ ions in many $AA_3'B_4O_{12}$ quadruple perovskites, Cu preferentially occupies the distorted octahedral $Mn_B$ sites in the present system. Consequently, Cu substitution directly perturbs the $MnO_6$ octahedral network responsible for orbital ordering and magnetic exchange interactions. Local modifications of Mn-O-Mn bond angles and orbital overlap are therefore expected even though the average crystallographic symmetry remains monoclinic $I2/m$. Importantly, the PDF results indicate that the primary effect of Cu substitution is the introduction of local structural disorder rather than a complete long-range crystallographic transformation. This interpretation is also consistent with the relatively large $R_{int}$ values obtained from the single-crystal refinements, which suggest crystallographic domain formation, local strain, and nanoscale structural inhomogeneity in the Cu-doped samples. The enhanced damping of the PDF oscillations at larger *r* values further supports a reduction in medium-range structural coherence with increasing Cu concentration.

The local structural disorder revealed by the PDF analysis is likely strongly coupled to the magnetic properties of $BiMn_{7-x}Cu_xO_{12}$. In quadruple perovskite manganites, magnetic exchange interactions are highly sensitive to local bond geometry, octahedral distortions, and orbital ordering. Therefore, even subtle local structural changes induced by Cu substitution can substantially modify the balance between ferromagnetic and antiferromagnetic interactions, providing a natural explanation for the pronounced magnetic irreversibility, field-dependent hysteresis, and unsaturated magnetization observed experimentally. To further investigate the relationship between local structural disorder and magnetic behavior, detailed magnetic measurements were performed on the synthesized $BiMn_{7-x}Cu_xO_{12}$ samples.

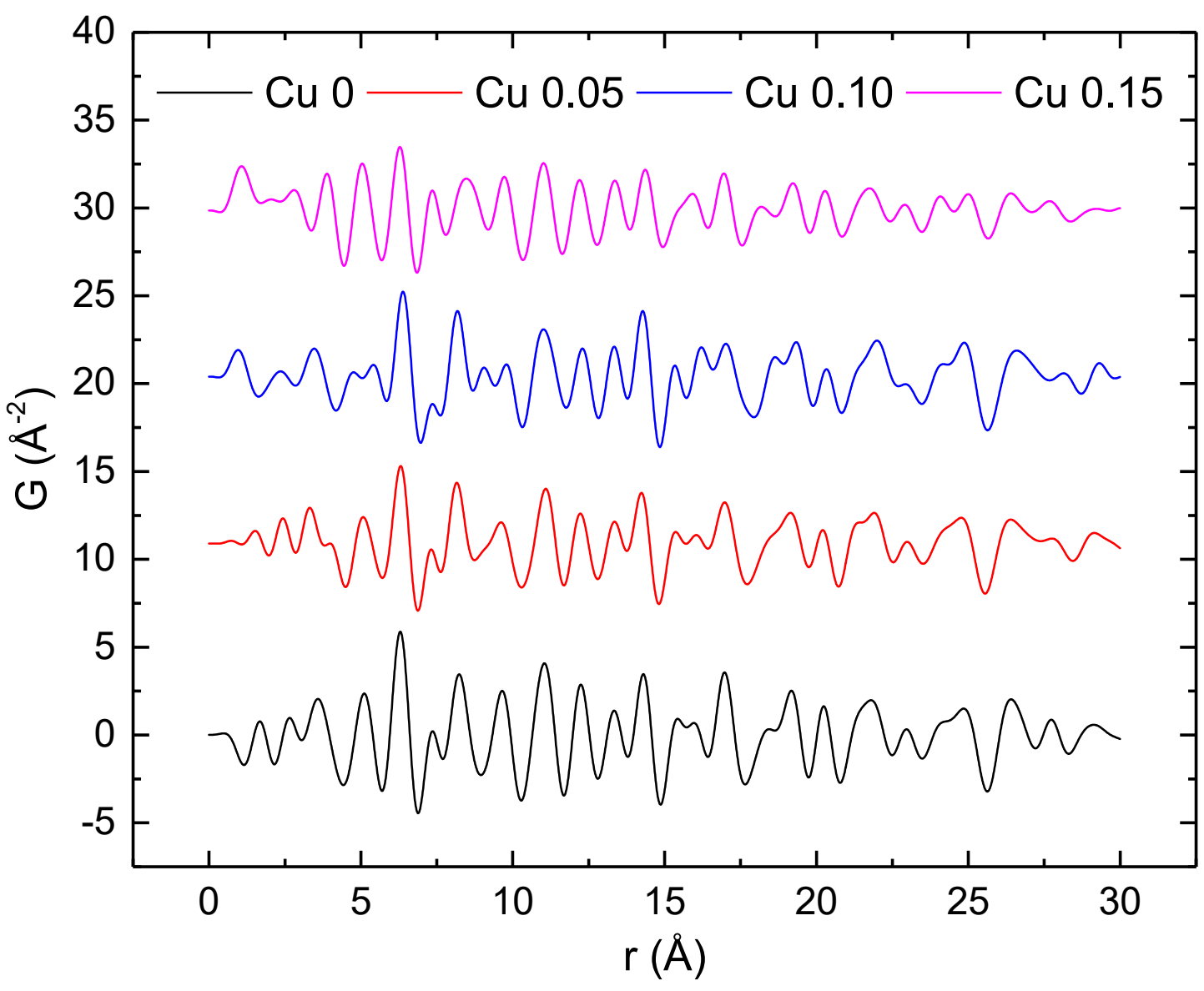

**Figure 3. Pair distribution function *G*(*r*) of $BiMn_{7-x}Cu_xO_{12}$ at different Cu concentrations.** The overall similarity in peak positions indicates that the local quadruple-perovskite framework is largely preserved upon Cu substitution, while subtle changes in peak intensity and damping reflect local disorder and modified Mn/Cu-O bonding environments.

The local structural disorder suggested by the SCXRD refinements and PDF analysis was further investigated using atomic-resolution HAADF-STEM imaging. As shown in **Figure 4**, the STEM images reveal a well-ordered quadruple-perovskite lattice over extended regions, confirming that the long-range framework remains intact after Cu substitution. The brightest atomic columns primarily correspond to Bi-containing sites because of the strong atomic-number contrast in HAADF imaging, whereas the weaker-intensity columns arise from the Mn/Cu and O sublattices. Although STEM cannot unambiguously distinguish the subtle monoclinic *I*2/*m* distortion from a pseudo-cubic projection, the comparison between the undoped and doped samples confirms that the overall framework is retained after Cu substitution. Elemental analysis further verifies the presence of Cu in the doped sample.

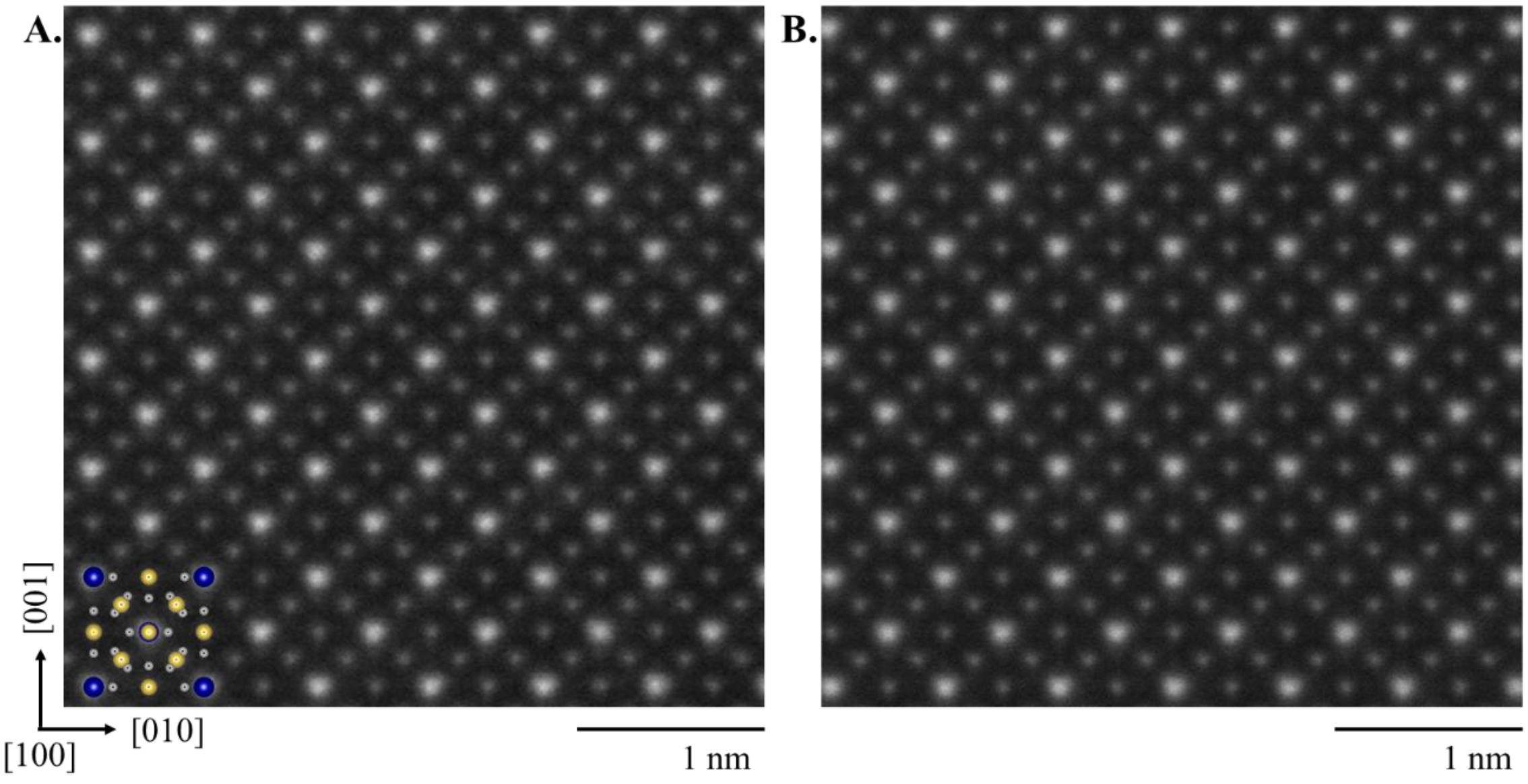

**Figure 4. Atomic-resolution HAADF-STEM images of A. $BiMn_7O_{12}$ and B. Cu-doped $BiMn_7O_{12}$.** Bright atomic columns correspond primarily to Bi-containing sites (Dark Blue) due to the strong atomic-number contrast in HAADF imaging, while weaker-intensity columns arise from Mn/Cu (Light Orange) and O (Grey Circle) sublattices.

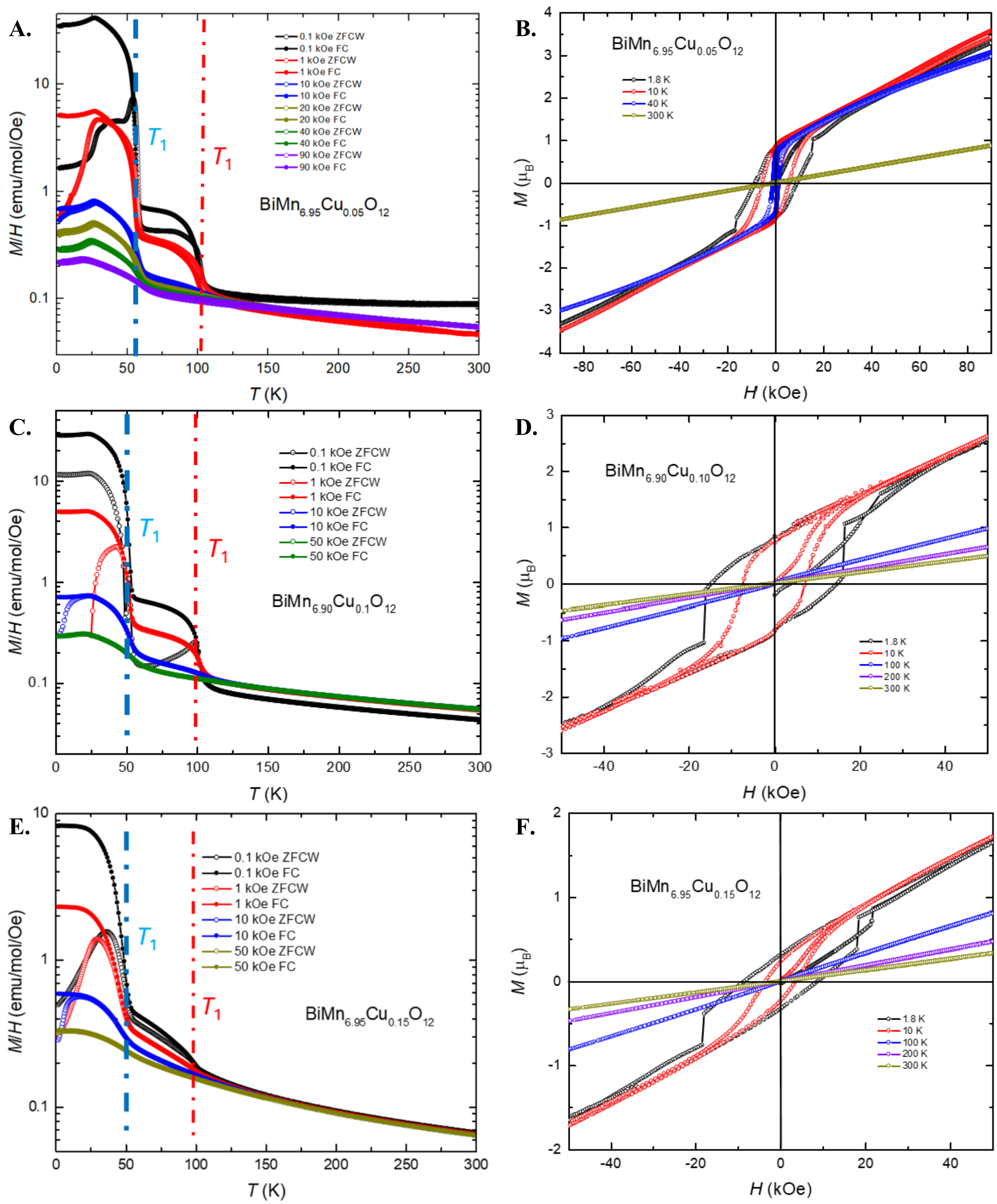

**Figure 5. Magnetic properties of $BiMn_{7-x}Cu_xO_{12}$ at different Cu concentrations. (A, C, E)** Temperature dependence of magnetic susceptibility measured under zero-field-cooled warming (ZFCW) and field-cooled (FC) conditions at selected magnetic fields for $x$ = 0.05, 0.10, and 0.15, plotted on a logarithmic scale. Two characteristic magnetic anomalies, $T_1$ and $T_2$, are indicated by dashed lines. **(B, D, F)** Magnetic hysteresis loops, $M(H)$, measured at selected temperatures for the corresponding compositions.

**Figure 5** presents the magnetic properties of $BiMn_{7-x}Cu_xO_{12}$ for $x$ = 0.05, 0.10, and 0.15. The temperature-dependent susceptibility measured under ZFCW and FC conditions reveals two characteristic magnetic anomalies for all three compositions. The higher-temperature anomaly occurs near $T_1 \sim$ 100-120 K, while the lower-temperature feature appears near $T_2 \sim$ 50-60 K. Above $T_1$, the susceptibility curves measured under different applied fields largely overlap, indicating an approximately linear magnetic response. This is consistent with the nearly linear $M(H)$ curves measured at high temperatures. Below $T_1$, a clear bifurcation between the ZFCW and FC curves develops at low magnetic fields, indicating the onset of magnetic irreversibility. This behavior becomes more pronounced below $T_2$, where susceptibility increases sharply and the low-field response becomes strongly history dependent. The bifurcation is gradually suppressed under larger applied fields and becomes much less obvious above approximately 20 kOe, suggesting that the low-temperature magnetic state is field sensitive and contains weakly pinned or easily reoriented magnetic moments.

The corresponding $M(H)$ curves further confirm the development of a low-temperature hysteretic magnetic state. For $x$ = 0.05, the 1.8 K and 10 K curves show large hysteresis and nonlinear field dependence, with the moment reaching approximately 3-3.6 $\mu_B$/f.u. at high field. For $x$ = 0.10, the hysteresis remains evident, but the maximum moment is reduced to about 2.5-2.7 $\mu_B$/f.u. For $x$ = 0.15, the hysteresis is further weakened, and the maximum moment decreases to approximately 1.5-1.7 $\mu_B$/f.u. Thus, increasing Cu concentration suppresses the field-induced magnetic moment and weakens the hysteretic response. Importantly, none of the $M(H)$ curves reach full saturation within the measured field range. Therefore, the low-temperature state should not be assigned as a simple ferromagnet. Instead, the coexistence of finite hysteresis, large field-induced moments, and unsaturated magnetization suggests a complex magnetic ground state, likely involving canted antiferromagnetism, ferrimagnetic components, or competing ferro- and antiferromagnetic exchange interactions.

The magnetic anomalies near $T_1$ and $T_2$ likely arise from changes in the balance of magnetic exchange interactions associated with local structural distortions, orbital rearrangements, or modulated lattice states. This interpretation is consistent with the published picture that Cu doping tunes competing Jahn-Teller and Bi lone-pair instabilities in $BiMn_7O_{12}$. In our samples, single-crystal diffraction and PDF analysis indicate that the average monoclinic framework is retained, while Cu substitution introduces local disorder and directly perturbs the Mn-O network. Therefore, the observed magnetic irreversibility and unsaturated hysteresis are best understood as consequences of local magnetostructural disorder rather than conventional long-range ferromagnetic ordering.

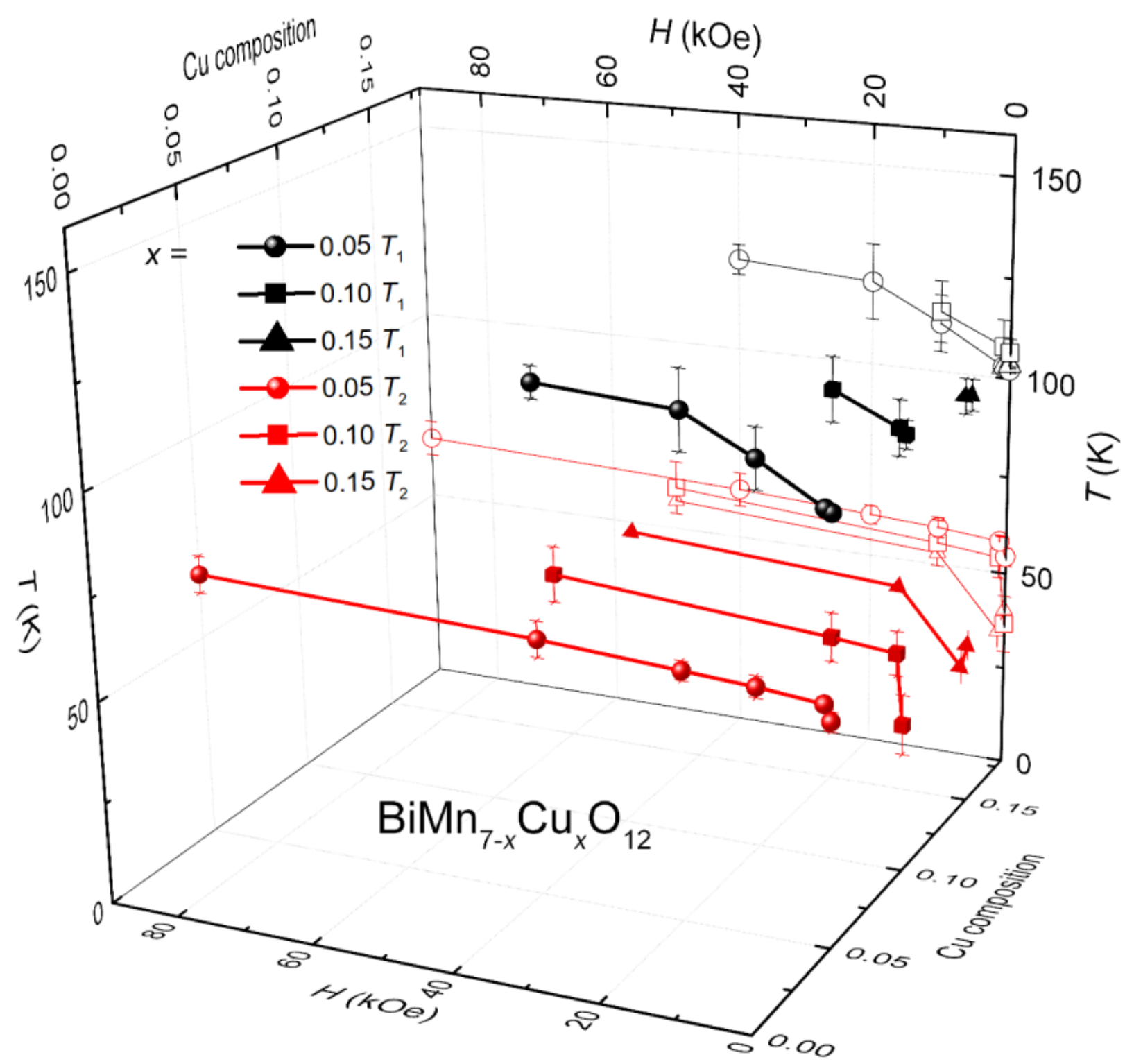

**Figure 6. Composition–magnetic field–temperature phase diagram of $BiMn_{7-x}Cu_xO_{12}$.** It summarizes the evolution of the characteristic magnetic temperatures $T_1$ and $T_2$ as functions of Cu concentration and applied magnetic field.

**Figure 6** summarizes the evolution of the characteristic magnetic temperatures $T_1$ and $T_2$ as functions of Cu concentration and applied magnetic field in $BiMn_{7-x}Cu_xO_{12}$. Both anomalies exhibit clear magnetic-field dependence, confirming their magnetic origin and indicating strong coupling between the magnetic state and external field. The two characteristic temperatures exhibit distinctly different composition dependences. The lower-temperature anomaly, $T_2$, systematically shifts to lower temperature with increasing Cu concentration, indicating progressive suppression of the low-temperature magnetic state upon Cu substitution. In contrast, $T_1$ remains comparatively stable and even shifts slightly upward under moderate magnetic fields. However, the magnetic signature associated with $T_1$ becomes increasingly broadened and fragile as the Cu concentration increases, especially for $x = 0.15$, where the anomaly is readily suppressed under applied magnetic field. Increasing magnetic field generally stabilizes the magnetic state and shifts both $T_1$ and $T_2$ toward higher temperature, although the effect is substantially stronger for $T_2$. This behavior is consistent with the field-dependent hysteresis observed in the $M(H)$ measurements and suggests that the low-temperature magnetic state contains a sizable field-responsive component. The contrasting evolution of $T_1$ and $T_2$ suggests that the two anomalies originate from different magnetic mechanisms or competing magnetic interactions. The $T_1$ feature likely corresponds to a more fragile magnetic instability associated with local structural/orbital fluctuations, whereas $T_2$ is associated with the development of the stronger low-temperature hysteretic magnetic state. The gradual suppression of $T_2$ and increasing fragility of $T_1$ with Cu substitution further indicate that Cu doping progressively perturbs the magnetic exchange network and destabilizes the competing magnetic ground states in $BiMn_{7-x}Cu_xO_{12}$.

To evaluate the Cu site preference and coloring problem in $BiMn_7O_{12}$, first-principles calculations were performed using the conventional $Bi_2Mn_{14}O_{24}$ cell containing 40 atoms, corresponding to two primitive cells. One Mn atom was substituted by Cu to model the dilute doping limit. Five crystallographically distinct Mn sites were considered: three square-planar A' sites, $Mn_{A1}$, $Mn_{A2}$, and $Mn_{A3}$, and two octahedral B sites, $Mn_{B1}$ and $Mn_{B2}$. The calculated relative energies, converted to eV per Cu, are summarized below.

| **Model** | **ΔE = ($E_{A'}$-$E_B$)** | **3.8% Cu ΔE = ($E_{A'}$-$E_B$)** | **Energetic Preference** |
|---|---|---|---|
| FM, relaxed | −174 meV/Cu | −6.6 meV | A' site |
| FM, fixed structure | −842 meV/Cu | −32.0 meV | A' site |
| FM + (U), fixed structure | −611 meV/Cu | −23.2 meV | A' site |

For the fully relaxed FM calculations, the lowest-energy A'-site configuration is Cu@$Mn_{A2}$, whereas the lowest-energy B-site configuration is Cu@$Mn_{B2}$. The A'-site configuration is lower by 0.174 eV/100%Cu. With the highest Cu loading composition, $x = 0.15$, the total energy difference is around 6.6 meV for 3.8% doping according to the experimental data. In the fixed-structure FM calculations, Cu@$Mn_{A1}$ is the most stable configuration and is lower than the best B-site configuration by 0.842 eV/Cu. When an effective Hubbard U = 4 eV is applied to Mn $3d$ orbitals, the A'-site preference remains, with Cu@$Mn_{A1}$ lower than the best B-site configuration by 0.611 eV/Cu. These calculations therefore indicate that, under the tested FM configurations and computational settings, the square-planar A' site is energetically favored for Cu substitution. This trend is consistent with the conventional crystal-chemical expectation that Jahn-Teller-active $Cu^{2+}$ often stabilizes square-planar coordination. However, this calculated preference differs from the single-crystal X-ray refinement results, where Cu occupancy is more strongly associated with the B-site electron density. This apparent discrepancy suggests that Cu site preference in $BiMn_{7-x}Cu_xO_{12}$ may be highly sensitive to structural relaxation, magnetic configuration, local disorder, oxygen stoichiometry, and high-pressure synthesis conditions.

Because several crystallographically distinct Mn sites exist within both the A' and B sublattices, the Cu substitution problem is also a configurational "coloring" problem: different ways of placing Cu on symmetry-inequivalent Mn sites generate different local bonding environments and total energies. Among the A' sites, the preferred site changes from $Mn_{A2}$ in the relaxed FM calculation to $Mn_{A1}$ in the fixed-structure and DFT+U calculations, showing that local relaxation and correlation effects influence the site hierarchy. Among the B sites, $Mn_{B1}$ and $Mn_{B2}$ are nearly degenerate, indicating that the two octahedral B sites have similar energetic stability for Cu incorporation. The calculations support a strong energetic competition among inequivalent Cu-coloring configurations. While DFT favors A'-site occupation under the tested assumptions, the experimental SCXRD results favor B-site occupancy. This contrast implies that the experimentally observed Cu distribution may not be controlled by simple equilibrium site energetics alone, but may also reflect kinetic trapping, local strain, domain formation, or synthesis-dependent metastability in this high-pressure quadruple perovskite.

## Conclusion

We investigated the structural and magnetic evolution of lightly Cu-doped $BiMn_7O_{12}$ using single-crystal X-ray diffraction, PDF analysis, STEM, magnetic measurements, and first-principles calculations. All investigated compositions retain an average monoclinic $I2/m$ structure, while Cu substitution progressively suppresses the monoclinic distortion and reduces the unit-cell volume. PDF analysis reveals increasing local structural disorder and reduced medium-range coherence upon Cu incorporation without a long-range symmetry-breaking transition. Single-crystal refinements indicate enhanced electron density at the octahedral Mn B sites, suggesting that Cu directly perturbs the Jahn-Teller-active $MnO_6$ network. In contrast, first-principles calculations favor Cu occupation at the square-planar A' sites within the tested FM-based computational models, revealing strong competition among local bonding environments and possible synthesis-dependent metastability. Magnetic measurements reveal two characteristic anomalies near $T_1$ and $T_2$, together with strong magnetic irreversibility and unsaturated magnetization. Increasing Cu concentration suppresses the low-temperature magnetic state and weakens the uncompensated magnetic component. These results demonstrate that even subtle local structural perturbations strongly influence the coupled orbital and magnetic instabilities in Cu-doped $BiMn_7O_{12}$, highlighting the importance of disorder-driven magnetostructural coupling in quadruple perovskite manganites.

## Supporting Information

Total scattering structure function $F(Q)$ and corresponding pair distribution function $G(r)$ of Cu-doped $BiMn_{7-x}Cu_xO_{12}$ (x = 0, 0.05, 0.10, 0.15). (DOS)

## Acknowledgments

C.P. at Michigan State University were supported by NSF-DMR-2422361. J. L. at the University of Michigan was supported by NSF-DMR-2422362. M.X. at Michigan State University were supported by the U.S. Department of Energy (DOE), Division of Basic Energy Sciences under Contract DE-SC0023648. W.X. at Michigan State University were supported by U.S. Department of Energy (DOE), Division of Basic Energy Sciences under Contract DE-SC0024943.

## Conflict of Interest

The authors declare no conflict of interest.

## Supplementary Information

# Site Preferences and "Coloring Problem" in Cu-doped $BiMn_7O_{12}$ Quadruple Perovskite

*Cheng Peng[1&], Mingyu Xu[1&], Yang Zhang[2], Ismail El Baggari[2,3], Jie Li[4], Weiwei Xie[1*]*

1. Department of Chemistry, Michigan State University, East Lansing, MI 48824 USA
2. The Rowland Institute at Harvard, Harvard University, Cambridge, MA 02138 USA
3. Department of Physics, University of British Columbia, Vancouver, BC V6T 1Z4 Canada
4. Department of Earth and Environmental Sciences, University of Michigan, Ann Arbor, MI 48109 USA

[*]Corresponding author: Dr. Weiwei Xie (xieweiwe@msu.edu)

[&]equally contributed

### Table of Contents

**Figure S1.** Total scattering structure function $F(Q)$ and corresponding pair distribution function $G(r)$ of $BiMn_7O_{12}$.

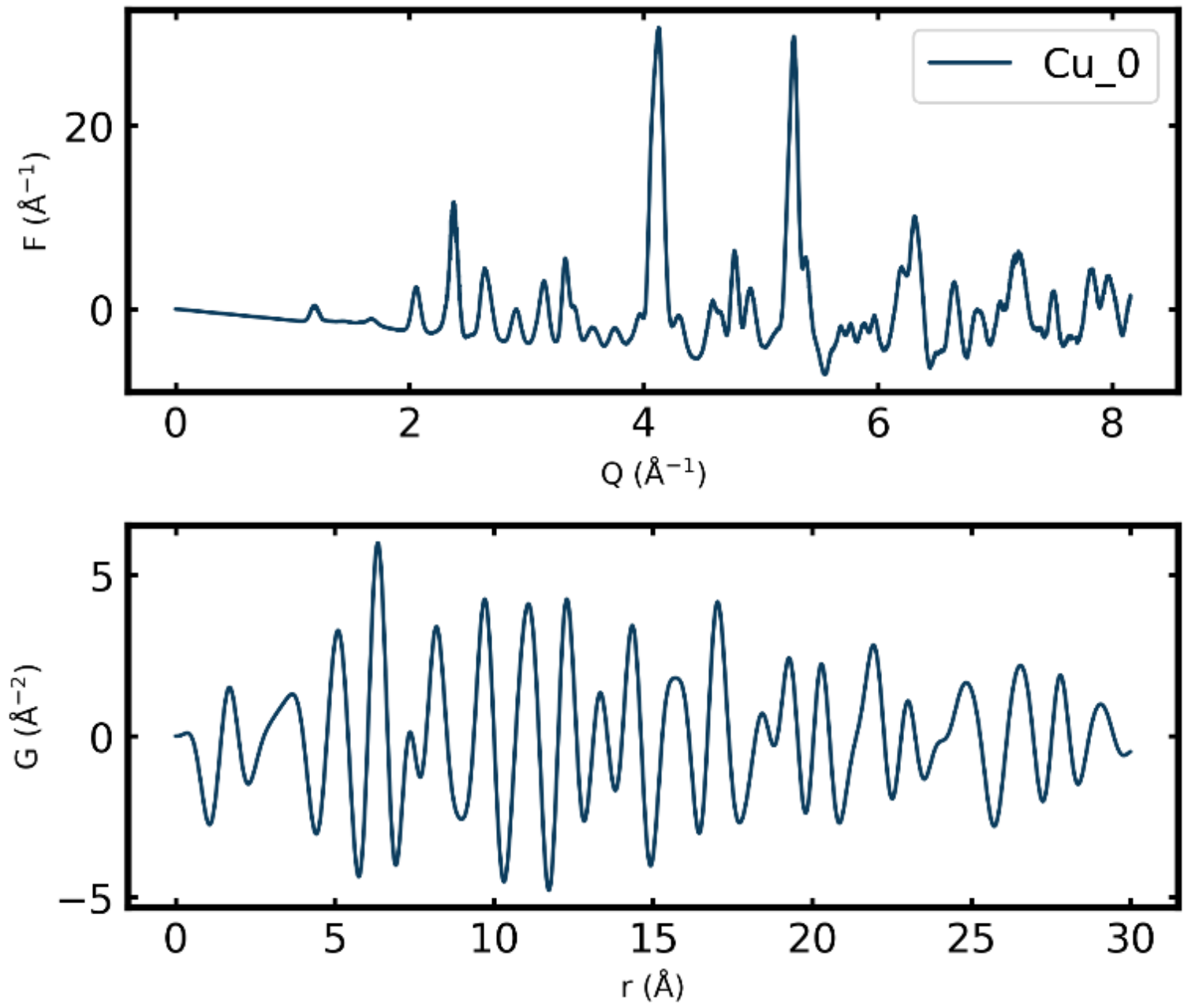

**Figure S2.** Total scattering structure function $F(Q)$ and corresponding pair distribution function $G(r)$ of $BiMn_{6.95}Cu_{0.05}O_{12}$.

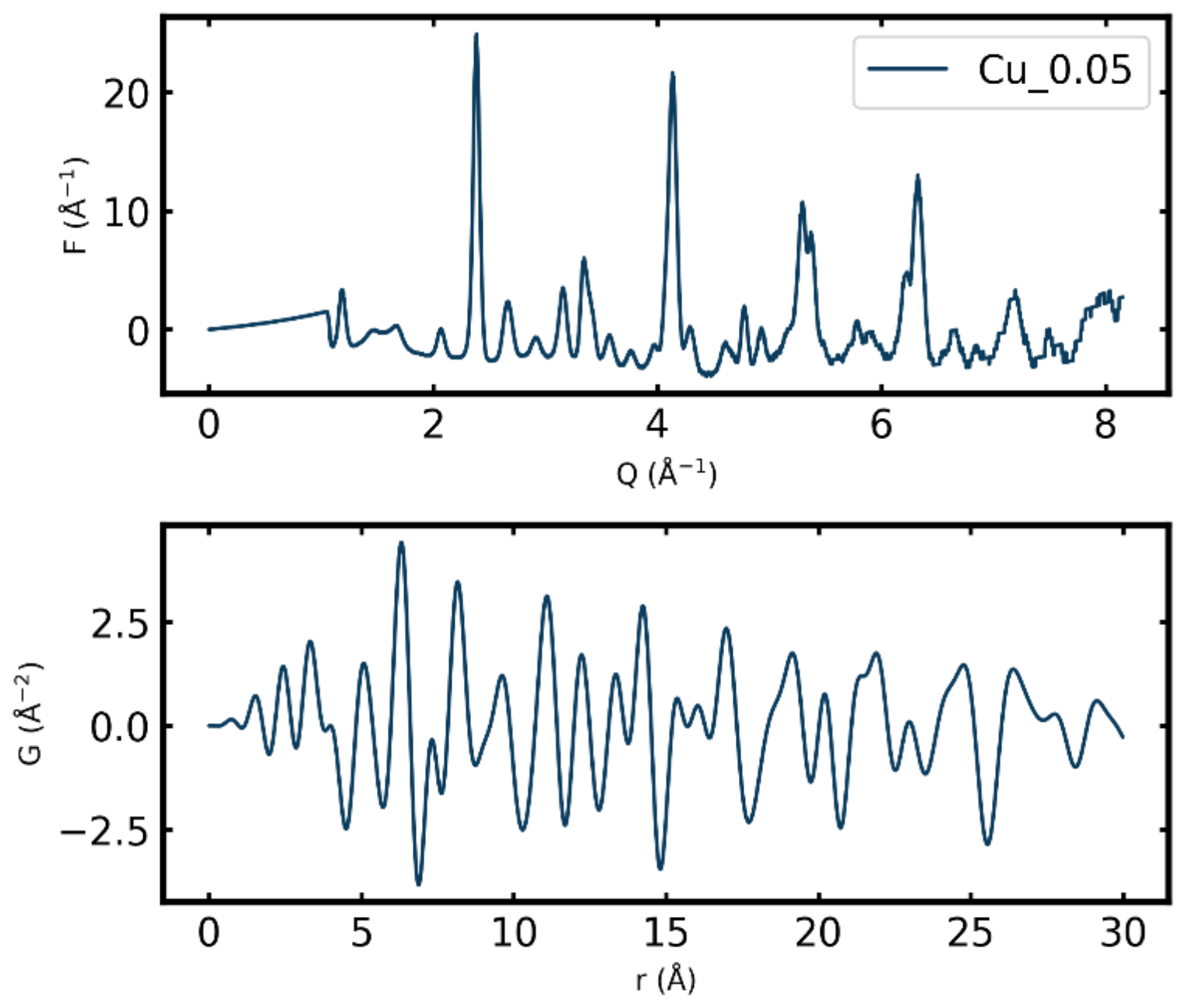

**Figure S3.** Total scattering structure function $F(Q)$ and corresponding pair distribution function $G(r)$ of $BiMn_{6.90}Cu_{0.10}O_{12}$.

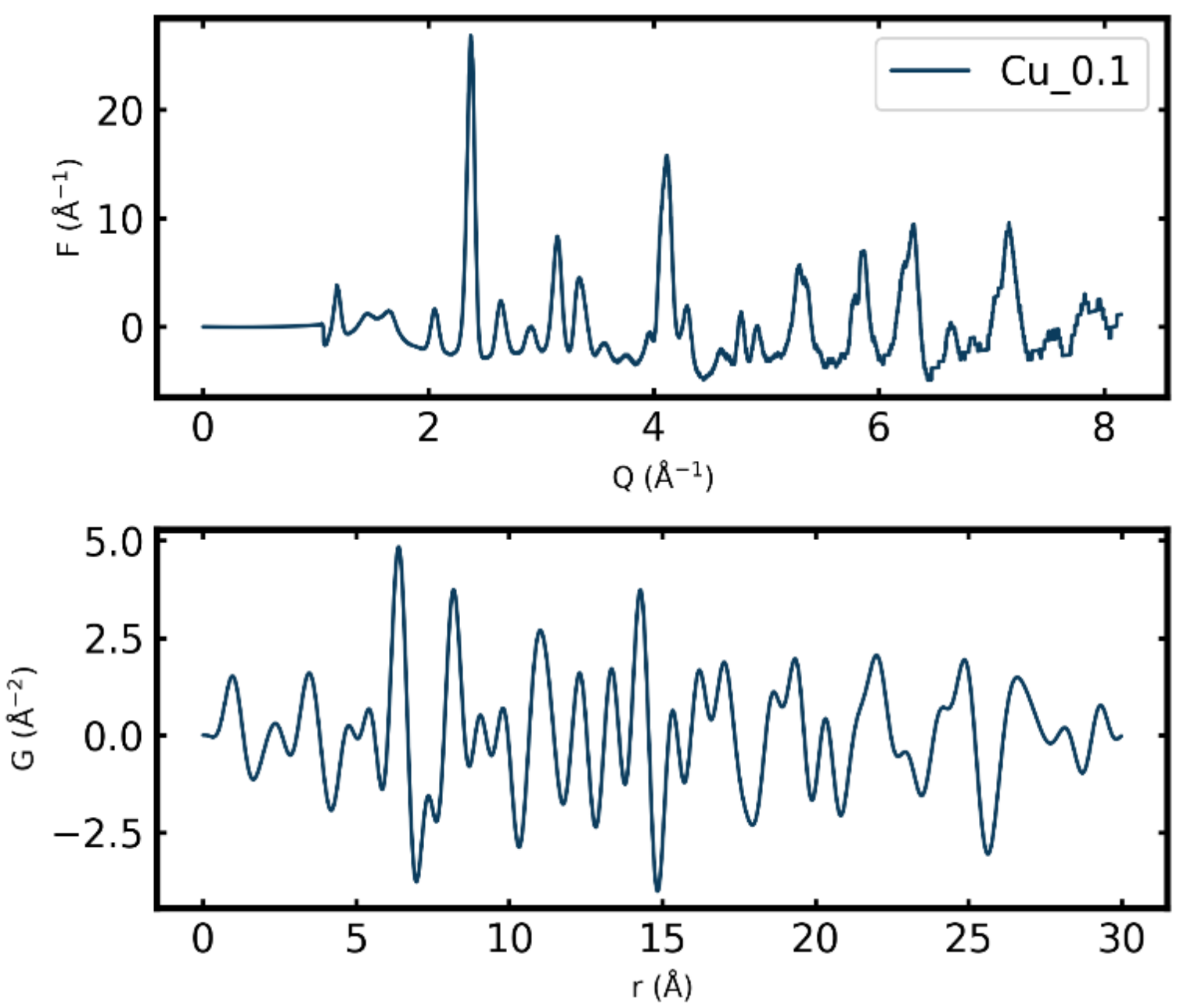

**Figure S4.** Total scattering structure function $F(Q)$ and corresponding pair distribution function $G(r)$ of $BiMn_{6.85}Cu_{0.15}O_{12}$.

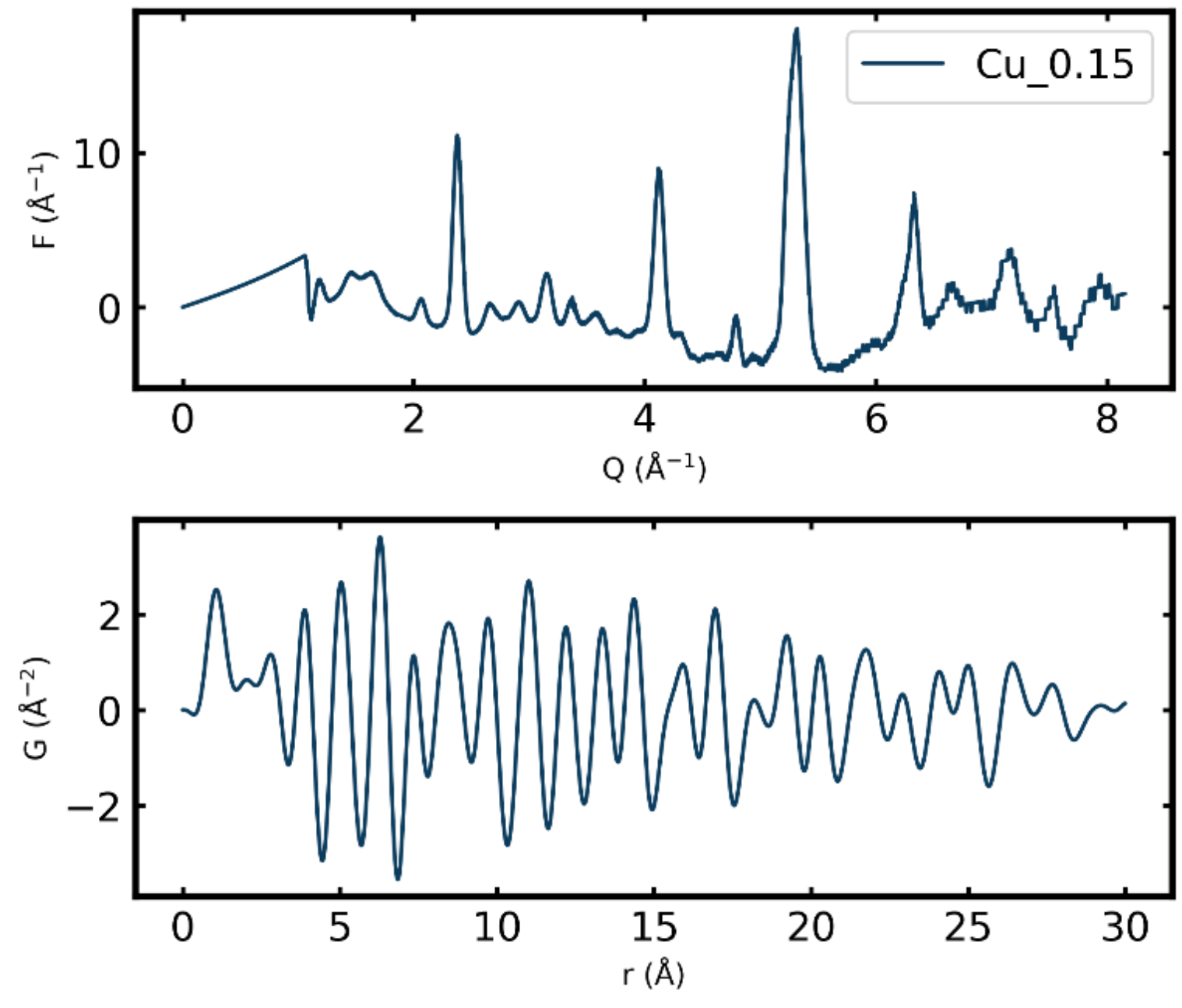